\documentclass[aps,pra,10pt,twocolumn,superscriptaddress]{revtex4}
\usepackage{amsmath}
\usepackage{amssymb}
\usepackage{bbm}
\usepackage{graphicx}
\usepackage{framed}
\usepackage{color}
\usepackage{hyperref}
\usepackage{epstopdf}
\usepackage{dsfont}
\usepackage{amssymb}
\usepackage{amsmath}
\usepackage{amsthm}
\usepackage{wrapfig}
\usepackage{relsize}
\usepackage{bm}
\usepackage[latin1]{inputenc}
\usepackage{enumitem}

\newcommand{\Tr}{{\rm Tr}}
\newcommand{\red}[1]{{\color{red}}}

\def\Tr{\hbox{Tr}} 
\usepackage{pifont}

\newcommand{\ket}[1]{\vert#1\rangle}
\newcommand{\bra}[1]{\langle#1\vert}

\newcommand{\ba}{\begin{eqnarray}}
\newcommand{\ea}{\end{eqnarray}}

\begin{document}

\title{Geometrical bounds on irreversibility in open quantum systems} 

\author{Luca Mancino}
\email{luca.mancino@uniroma3.it}
\affiliation{Dipartimento di Scienze, Universit\`{a} degli Studi Roma Tre, Via della Vasca Navale 84, 00146, Rome, Italy}

\author{Vasco Cavina}
\affiliation{NEST, Scuola Normale Superiore and Istituto Nanoscienze-CNR, Piazza dei Cavalieri 7, I-56126, Pisa, Italy.}

\author{Antonella De Pasquale}
\affiliation{NEST, Scuola Normale Superiore and Istituto Nanoscienze-CNR, Piazza dei Cavalieri 7, I-56126, Pisa, Italy.}
\affiliation{Dipartimento di Fisica, Universit\`{a} di Firenze, Via G. Sansone 1, I-50019 Sesto Fiorentino (FI), Italy.}

\author{Marco Sbroscia}
\affiliation{Dipartimento di Scienze, Universit\`{a} degli Studi Roma Tre, Via della Vasca Navale 84, 00146, Rome, Italy}

\author{Robert I. Booth}
\affiliation{Institut de Physique, Sorbonne Universit\'{e}, 4 Place Jussieu, 75005, Paris, France}
\affiliation{Dipartimento di Scienze, Universit\`{a} degli Studi Roma Tre, Via della Vasca Navale 84, 00146, Rome, Italy}

\author{Emanuele Roccia}
\affiliation{Dipartimento di Scienze, Universit\`{a} degli Studi Roma Tre, Via della Vasca Navale 84, 00146, Rome, Italy}

\author{Ilaria Gianani}
\affiliation{Dipartimento di Scienze, Universit\`{a} degli Studi Roma Tre, Via della Vasca Navale 84, 00146, Rome, Italy}

\author{Vittorio Giovannetti}
\affiliation{NEST, Scuola Normale Superiore and Istituto Nanoscienze-CNR, Piazza dei Cavalieri 7, I-56126, Pisa, Italy.}

\author{Marco Barbieri}
\affiliation{Dipartimento di Scienze, Universit\`{a} degli Studi Roma Tre, Via della Vasca Navale 84, 00146, Rome, Italy}
\affiliation{Istituto Nazionale di Ottica - CNR, Largo Enrico Fermi 6, 50125, Florence, Italy}

\begin{abstract} 
Clausius inequality has deep implications for reversibility and the arrow of time. Quantum theory is able to extend this result for closed systems by inspecting the trajectory of the density matrix on its manifold. Here we show that this approach can provide an upper and lower bound to the irreversible entropy production for open quantum systems as well. These provide insights on the thermodynamics of the information erasure. Limits of the applicability of our bounds are discussed, and demonstrated in a quantum photonic simulator.
\end{abstract}

\maketitle

\textit{Introduction.} 
Irreversibility in physical processes is strictly related to the idea of energy dissipation.
This concept is one of the cornerstones 
of thermodynamics since it came to the stage
in the second half
of the nineteenth century.
The second law of thermodynamics allows to give a quantitative characterization
to the interplay between the irreversibility and the
exchange of energy, introducing a state function, the entropy, that is always non decreasing in 
macroscopical processes \cite{Clausius1867}. 
The generality of this principle is a 
key feature in establishing
thermodynamics as one of the 
fundamental branches of classical theories
and plays a  role
also in the connections between physics and
information science \cite{Landauer1961, Goold2016, Plenio2001}.
A better understanding of the implication of the second law at the quantum level
is one of the main issues of the actual research in the field of quantum thermodynamics
\cite{Horodecki2011,Lostaglio2015,Cwiklinski2015, Vinjanampathy2016} and is fundamental in the construction
of a solid theoretical ground for a wide class of applications such as computation \cite{Bennet2003},
metrology \cite{Yuasa17, DePalma17, Correa15, dechiara}, quantum control \cite{Feldmann2010, Cavina2017bis, Cavina2016} and 
quantum thermal engines \cite{DelCampo2014, Alicki1979, Cavina2017}.

\begin{figure}[b]
\includegraphics[width=\columnwidth]{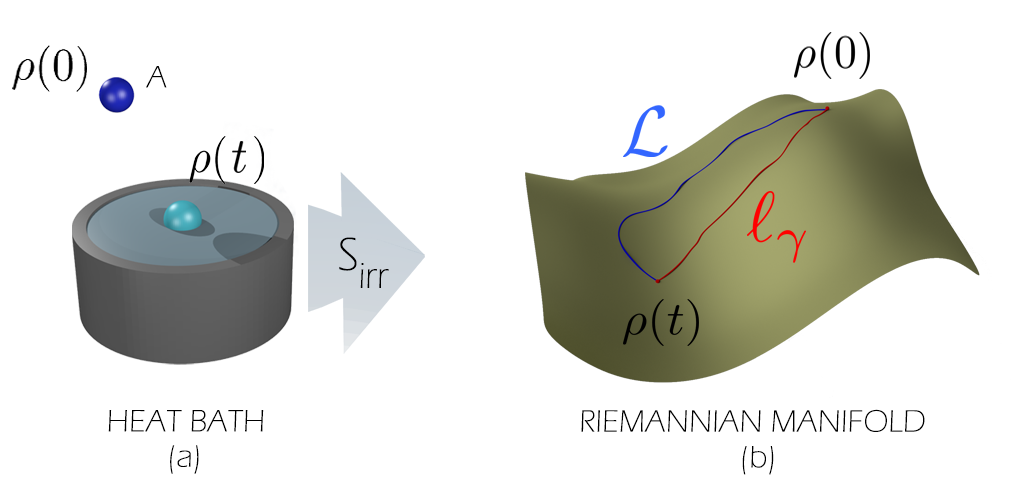}
\caption{Principal features of our scheme: (a) the qubit is initialized in a given state, then it interacts with a bosonic reservoir at temperature $T$; such irreversible process produces irreversible entropy; (b) following the evolution on the state space allow for information-geometric considerations delivering a lower bound on the produced entropy.}
\label{Protocol}
\end{figure}

In this work we examine the thermalization process of a quantum system $\cal A$ in 
contact with a reservoir, as pictorially represented in Fig.~\ref{Protocol}. For this 
significant example of out-of-equilibrium evolution we manage to bound the irreversible
entropy production both from above and below, facing
the problem within an informational-geometric setting \cite{Ruppeiner1979, Salamon1983}.
While the upper bound obtained here is completely
general, the lower bound holds only under 
certain conditions, satisfied for example 
in the single qubit scenario.

\begin{figure*}[t]
\includegraphics[width=0.9\textwidth]{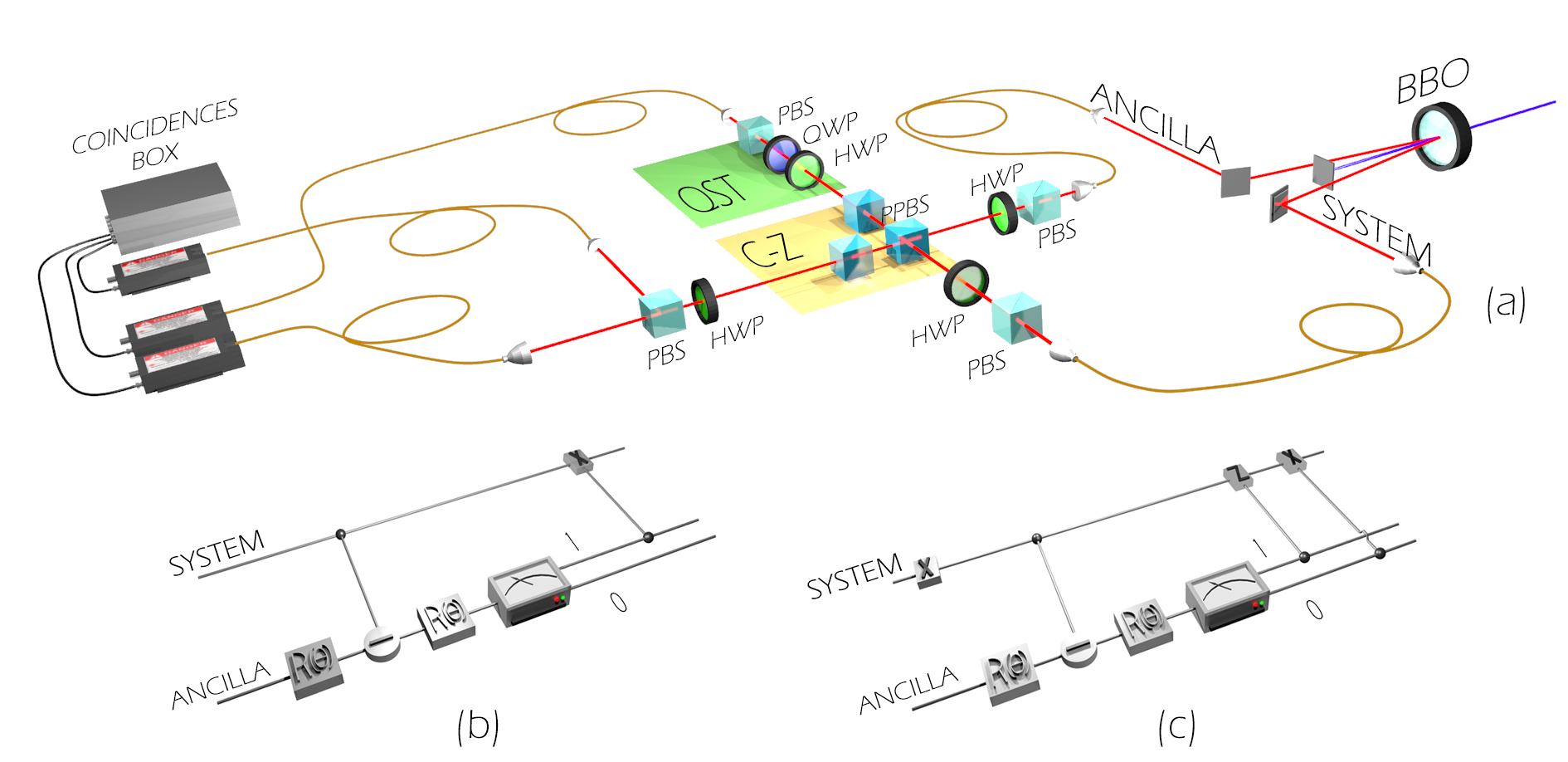}
\caption{Quantum simulation via quantum logic gates. a) Experimental apparatus. The system and the ancilla photons are generated via a Spontaneous Parametric Down Conversion (SPDC) process through a $\beta$ Barium-Borate crystal (Type I) pumped with a Continuous Wave (CW) laser. List of abbreviations: PBS=Polarizing Beam Splitter, PPBS=Partially Polarizing Beam Splitter, HWP=Half Wave Plate, QWP=Quarter Wave Plate, QST=Quantum State Tomography. b) Quantum circuit of the AD channel, c) Quantum circuit of the IAD channel. }
\label{Setup}
\end{figure*}

We will treat the Hilbert space of the system $\cal A$ as a Riemannian manifold. This allows us to establish relations between the irreversible entropy production and the geodesic distances corresponding 
to metrics that are
contractive under complete positive and trace preserving (CPTP) maps.
Following a similar approach, 
S. Deffner and E. Lutz~\cite{Deffner2010} determined a lower bound  to the irreversible entropy production for a closed
quantum system driven by an external field,
in terms of the Bures
length. 
However, it results that the latter is not the only 
contractive Riemannian metric on the system Hilbert space, as there exists an infinite 
family of such metrics as characterized by the  Morozova, \u{C}encov and Petz
theorem~\cite{Morozova1991, Petz1996}. Within this {\it mare magnum}, the only 
Riemannian metrics whose geodesics are analytically known are the Bures and the 
Wigner-Yanase one. Quite recently, Pires {\it et. al}~\cite{Pires2016} proved that
the Bures metrics for the case of any single qubit unitary dynamics, effectively provides a tighter bound on the transformation speed at least when compared to the Wigner-Yanase one; these transformations 
encompass those considered in~\cite{Deffner2010}. In general, 
establishing which metrics returns the smallest geodesic distance between states
is still an open question \cite{Pires2016, Luo2004}. In our manuscript we compare the two above mentioned 
metrics  for the case of a qubit thermalisation process described by the so-called generalized amplitude damping map and show that in this case the Wigner-Yanase one provides the sharper bounds to the
irreversible entropy production. Our results are supported by experimental evidence, gained by  simulating the
thermalization of a qubit in contact with a bosonic bath, using quantum photonics \cite{Mancino2017, Tham2017}. 
We have compared the performances of two states, a coherent and non-coherent one
in the canonical basis of $\cal A$, and find that only the former is able to
efficiently distinguish the performances of the two metrics.  
On a more practical ground, our bounds can be simply computed by performing the tomography of the state of the system at fixed time instants. In some cases, this represents a less demanding task than directly measuring thermodynamic functionals due to the non-linear expressions of the latter.


\begin{figure*}[t!]
\includegraphics[width=1\textwidth]{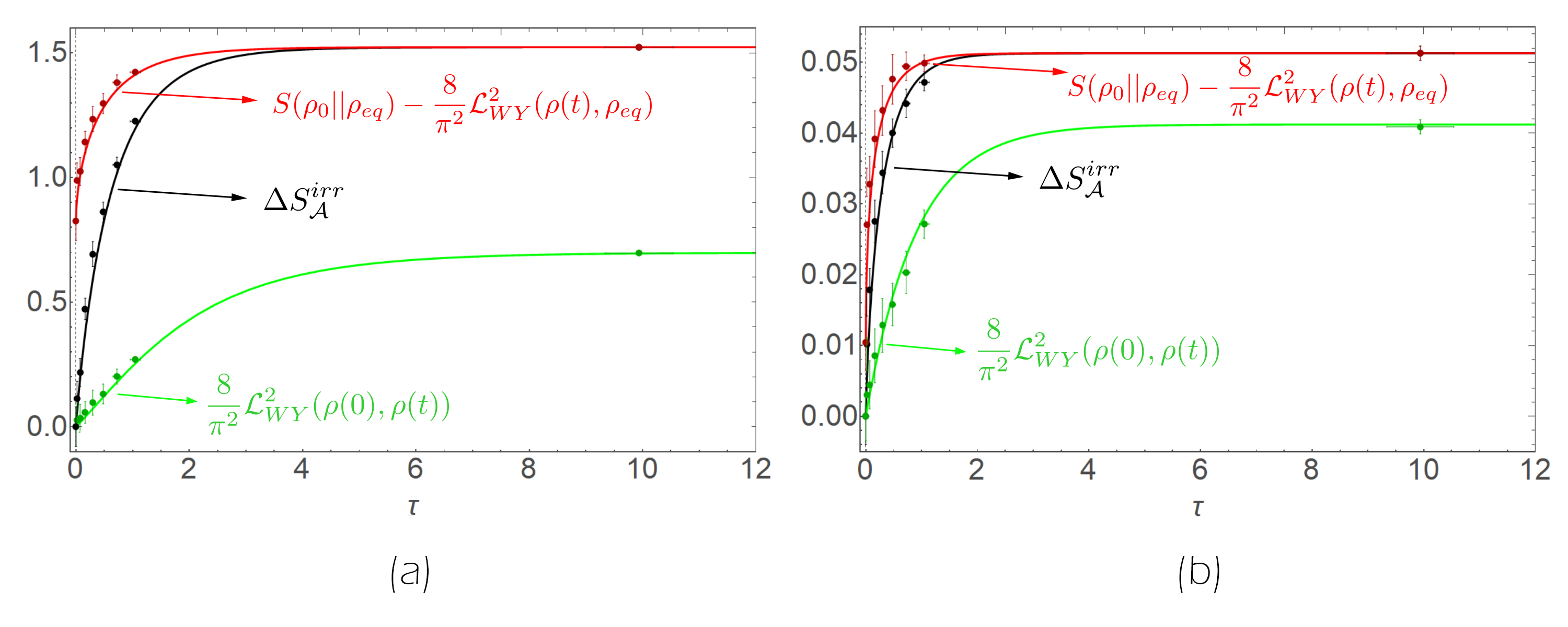}
\caption{Bounds to the irreversible entropy production. The initial state is $\vert D \rangle=1/\sqrt{2} (\vert H \rangle + \vert V \rangle)$ in panel (a) and $\vert V \rangle$ in panel (b).  The solid lines are theoretical predictions; experimental points have been obtained from the density matrices retrieved through Quantum State Tomography. Vertical errors have been obtained through a Monte Carlo routine that takes into account the Poissonian statistics of the measured counts, while horizontal errors depend on the uncertanty on the rotation $R(\theta)$ \cite{Mancino2017}.}
\label{Results}
\end{figure*}

{\it Geometrical bounds for irreversible entropy production.} Let us consider an arbitrary system $\cal A$ weakly coupled with a thermal bath at temperature $T$. Without loss of generality let us fix as $t=0$ the initial time $t$ of the process, and let $\rho_0$ be the associated density matrix of the system. If we keep the system Hamiltonian $H$ unchanged, the system will thermalize with the reservoir, thus asymptotically reaching the canonical equilibrium state $\rho_{eq}=e^{-\beta H}/ {\cal Z_{\beta}}$, where $\beta=1/k_B T$, $k_B=1$ is the Boltzmann constant and ${\cal Z}_\beta$ is the partition function. This represents a prototypical example of an irreversible thermodynamic evolution, which for instance finds applications also in the context of quantum thermometry~\cite{Correa15, DePalma17, Yuasa17, Mancino2017, Tham2017, Jevtic15, Genoni17}.  From a mathematical point of view, this process can be described by a two-parameters family of quantum channels $\{\Phi^{T}_t\}$ such that for all $t$ associates the density matrix $\rho(t):=\Phi^{T}_t[\rho_0]$ of the system to the initial state $\rho_0$. This maps admits a unique stationary state corresponding to the (unique) equilibrium state of the map, i.e. $\Phi^{T}_t\ (\rho_{eq}) =\rho_{eq}$. The thermodynamically irreversible component of the entropy variation $\Delta S_{\cal A}^{irr}$ produced due to evolution from $t=0$ to a generic time $t$ is given by
\begin{equation}
\Delta S_{\cal A}^{irr}:=\Delta S_{\cal A} - \frac{\Delta Q}{T},
\end{equation} 
where the first term on the right hand side is the entropy variation of the system $\Delta S_{\cal A}=S_{\cal A}(\rho(t)) - S_{\cal A}(\rho_0)$, being $S_{\cal A}(\rho)=- \Tr (\rho \ln \rho)$ the 
Von Neumann entropy of the density matrix $\rho$, and second term  is the heat $\Delta Q = \Tr[H \rho(t)] - \Tr[H \rho_0]$ absorbed by the system and corresponds to thermodynamically reversible contribution to the entropy production. The Clausius inequality, $\Delta S_{\cal A}^{irr} \geq 0$, provides a process-independent lower bound to $\Delta S_{\cal A}^{irr}$. 
In this manuscript we will write both a sharper lower bound and also an upper bound to this thermodynamic functional, by recasting it as \cite{Breuer2002,DeffnerOpen2011}
\begin{equation}
\label{sigma}  
\Delta S_{\cal A}^{irr} (t)=S(\rho_0|| \rho_{eq}) -S(\rho(t)||\rho_{eq}),
\end{equation}
where $S(\rho_1||\rho_2)= \Tr(\rho_1 \ln \rho_1) - \Tr(\rho_1 \ln \rho_2)$ is the so-called relative entropy of $\rho_1$ to $\rho_{2}$. Notice that in this case the Clausius inequality can be easily proved from~\eqref{sigma}, by exploiting the monotonicity of the relative entropy under quantum channels, i.e completely positive trace preserving operations.
%
Our interest for this relation stems from 
an inequality of geometrical nature that links the quantum relative entropy with the unitarily invariant 
norms on the Hilbert space.  More precisely, calling $\mathcal{L}$ the distance induced on the Hilbert space by a given unitarily invariant norm we have  $S(\rho||\sigma) \geq 2 \mathcal{L}^2(\rho, \sigma) / {\mathcal{L}^2(e_{1,1},e_{2,2} )}$ where $e_{i,j}$ is the matrix with $i,j$ element equal to 1 and
all  other  elements 0 \cite{Audenaert2005}.
We require the distance $\mathcal{L}$ to be contractive under the action of any map $\Lambda$ \textit{i.e.} $\mathcal{L} (\Lambda (\rho), \Lambda (\sigma)) \leq \mathcal{L} (\rho, \sigma)$. The latter condition is crucial in order to interpret $\mathcal{L}^2$ as a measure of the distinguishability between $\rho$ and $\sigma$. This is satisfied by 
an infinite family of metrics explicitly characterized by the Morozova, \u{C}encov and Petz theorem
\cite{Morozova1991, Petz1996, Cencov1982}. It follows that each of such metrics will provide a different consistent bound to the relative entropy. The only cases in which an analytical expression for the geodesic distance is known \cite{Pires2016, Luo2004} are the Quantum Fisher information metric $\mathcal{L}_{QF}(\rho, \sigma) = \arccos\big[\Tr[\sqrt{\sqrt{\rho} \sigma \sqrt{\rho}}]\big]$
and the Wigner-Yanase metric,
$\mathcal{L}_{WY}(\rho, \sigma) = \arccos\big[\Tr[\sqrt{\rho} \sqrt{\sigma}]\big]$.
Notice that if $\rho$ and $\sigma$ commute, the two geodesic distances coincide. We can use these properties to obtain a sharper geometrical bound to the irreversible component of the entropy during a thermalization process as: 
\begin{equation}
\label{upper}  \Delta S_{\cal A}^{irr} (t) \leq S(\rho_0||\rho_{eq}) - \frac{8}{\pi^2} \max_{\{X=QF, WY\}}\mathcal{L}_{X}^2 (\rho(t), \rho_{eq}), 
\end{equation}
where we exploited that $\mathcal{L}^2_{X}(e_{1,1},e_{2,2}) = {\pi^2}/{4}$, $X=QF,WY$. This is the upper bound to $\Delta S_{\cal A}^{irr}$ we are looking for: the geometrical distance establishes how much entropy is dissipated during a generic transformation of a quantum open system. Notice that asymptotically, i.e. for $t\to\infty$, the inequality above becomes trivially strict, as both members reduce to $S(\rho_0||\rho_{eq})$ (see expression~\eqref{sigma}).

The same technique can be exploited to determine a lower bound for $\Delta S_{\cal A}^{irr }$. However in this case an additional step that restricts our analysis to a specific class of dynamical maps is required. In the specific, we say that a dynamical map $\Phi_t$ satisfies a (reverse) triangle inequality 
for the relative entropy if, taken an initial state $\rho_0$ and two instants of time $t_1$ and $t_2\geq t_1$, 
we have
\ba \label{tin}
S(\rho_0||\Phi_{t_2}(\rho_0)) \geq S(\rho_0||\Phi_{t_1}(\rho_0)) + S(\Phi_{t_1}(\rho_0)||\Phi_{t_2}(\rho_0)).
\ea
Let us remark that the relation above is not valid in general. For instance it can be shown to hold when $\Phi_{t}(\rho_0) = (1-\lambda_t) \rho_0 + \lambda_t\rho_{eq}$, being $\lambda_t$ an increasing function of time $t$ (see the Supplementary Material). A less trivial case is provided by the Generalized Amplitude Damping channel for qubit systems, specifically considered in the following part of the manuscript.
However if~\eqref{tin} is satisfied by the dynamical process $\Phi_t^T$ and we choose $t_1=t$ and $t_2 = \infty$, we find:

\begin{equation}
\label{lower} \Delta S_{\cal A}^{irr} (t) \geq  \frac{8}{\pi^2} \max_{\{X=QF, WY\}}\mathcal{L}_{X}^2 (\rho_0, \rho(t)).
\end{equation}

Equation (\ref{lower}) represents a tighter version of the Clausius inequality and
generalizes to the open system 
framework the results obtained in \cite{Deffner2010}.

From now on, we will focus on the specific case in which the reference system is a qubit in thermal contact with a bosonic reservoir. Its evolution can be formally described in terms of the Generalized Amplitude Damping (GAD) channel, $\rho(t)=\Phi_t^T(\rho_0)$, with Kraus operators: $E_1=\sqrt{p}(\vert 0 \rangle \langle 0 \vert + \sqrt{1-\eta_t} \vert 1 \rangle \langle 1 \vert)$, $E_2=\sqrt{p} \sqrt{\eta_t} \vert 0 \rangle \langle 1 \vert$, $E_3=\sqrt{1-p} (\sqrt{1-\eta_t} \vert 0 \rangle \langle 0 \vert + \vert 1 \rangle \langle 1 \vert)$, $E_4=\sqrt{1-p} \sqrt{\eta_t} \vert 1 \rangle \langle 0 \vert$, in 
the canonical basis $\{|0\rangle, |1\rangle\}$ corresponding respectively to the excited and the ground state of the system. 
Here $p$ and $\eta_t \in [0,1]$ are the time- and temperature-dependent probability and damping coefficient respectively, related to the average boson occupation number $\bar{N}=(\coth[1/(2T))]{-}1)/2$ for the bath and to the (dimensionless) time $t$ as $(1-2p)=(1-2\bar{N})^{-1}$ and $\eta_t=1-e^{-(1-2\bar{N})t}$ \cite{Jevtic15,Mancino2017}. As anticipated above, such thermalizing map can be shown to fulfill the triangle inequality \eqref{tin}, thus the geometrical lower bound \eqref{lower} can be enforced \cite{Supplemental}.
Notice that the GAD map can be realized by combing an Amplitude Damping (AD) map described by $E_1$ and $E_2$ with an Inverse Amplitude Damping (IAD) map described $E_3$ and $E_4$.


\begin{figure}[t]
\includegraphics[width=\columnwidth]{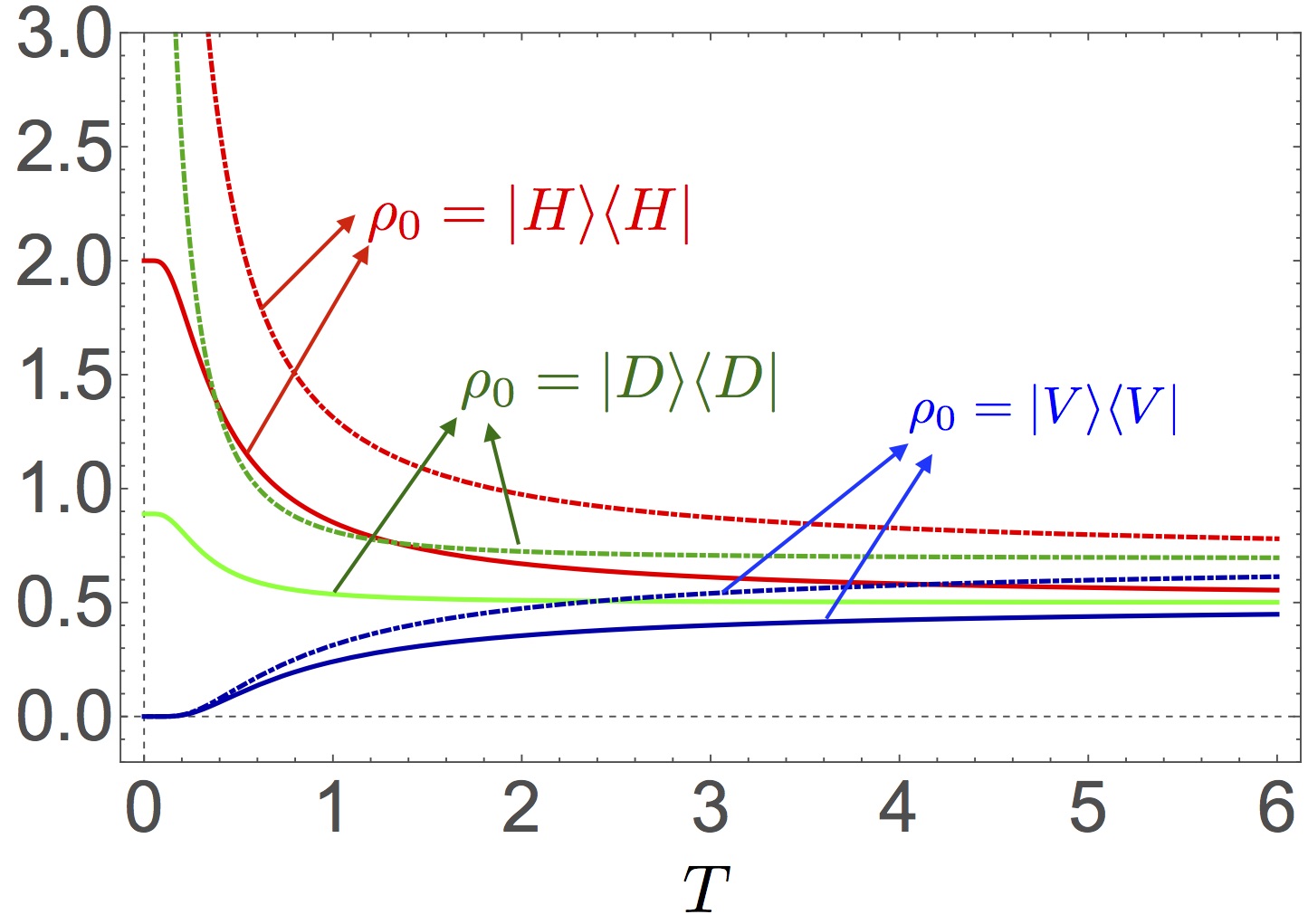}
\caption{(Color online) Asymptotic irreversible entropy production as a function of the temperature $T$ for the thermalization of the system $\cal{A}$. The dotted curves show the predictions associated to different pure states $\vert H \rangle$ (red), $\vert D \rangle$ (green) and  $\vert V \rangle$ (blue). The solid curves show the geometric lower bound \eqref{lower} in the thermalization limit for the same three cases, with the same color code, which provides a quite a good prediction for $ \Delta S_{\cal A}^{irr}$ for almost all values of $T$.}
\label{Fig4}
\end{figure}

\textit{Quantum photonic simulation.} We have simulated the thermalization dynamics of a qubit via a GAD channel, to test inequalities~\eqref{upper} and~\eqref{lower}. Our experiment is based on a quantum photonics logic gate, as a method to implement general quantum channels~\cite{7,Lu17} (other approaches still in framework of quantum photonics can be found in~\cite{cuevas1,cuevas2}). Our setup is shown in Fig.\ref{Setup} (a), and allows to realize the following steps: i) the canonical basis is encoded in the horizontal and vertical polarization states, i.e. $|0\rangle\equiv |H\rangle$ and $|1\rangle\equiv |V\rangle$. An ancilla qubit ${\cal B}$ is initialized in $\vert H \rangle$, while the target qubit ${\cal A}$ is prepared in a linear polarization state; ii)
the qubits, $\cal A$ and $\cal B$, interact through a Controlled-Sign (C-Z) gate, acquiring a $\pi$-phase on the $\vert V \rangle_{\cal A} \vert V \rangle_{\cal B}$ component only~\cite{8,9,10,11,12,13}; ${\cal B}$ undergoes the same rotation $R(\theta)$ before and after the C-Z, transmitted by Half Wave Plates (HWPs) set at an angle $\theta$: this controls the damping rate as $\eta=\sin^2 (4\theta)$; iii) the $H$/$V$ component of ${\cal B}$ is  measured and the outcomes control a Pauli transformation on $\cal A$. The arrengements in Figs.~\ref{Setup}(b) and~\ref{Setup}(c) implement the AD and IAD maps respectively;
iv) the density matrix $\rho(t)$ of the target $\cal A$ at time $t$  is determined collecting coincidence counts for the quantum state tomography in the two configurations, including the controlled-Pauli operations in post-processing. We then run the state reconstruction algorithm on data weighted with $p$ and $1-p$~\cite{Mancino2017}.

%
%
%

From the knowledge of full density matrix of the target qubit we have been able to determine the relevant quantities in the bounds~\eqref{upper} and \eqref{lower}, reported in Fig. \ref{Results} for $T=0.34$.  It can be numerically proved that the Wigner-Yanase metric, compared to the Bures one, provides a sharper bound to the irreversible entropy, for arbitrary choices of the initial state $\rho_0$ (the two metrics give the same result when $\rho_0$ is diagonal in the chosen basis). An experimental test for the state $\vert D \rangle=1/\sqrt{2} (\vert H \rangle + \vert V \rangle)$, analyzed in Fig.~\ref{Results}, is provided in the Supplementary Material. For this reason, in Fig.~\ref{Fig4} we report the experimental points associated to the Wigner-Yanase metric. The upper bound captures more closely the behaviour of $\Delta S_{\cal{A}}^{irr}$. Data closely follow the predictions with a small sistematic overestimation of the entropies. 

In the limit $t \rightarrow \infty$, the bound \eqref{lower} gives insights on the entropy produced for the erasure of qubit $\mathcal{A}$ by thermalization \cite{Lubkin1987,Vedral2002}. Our generalization allows to distinguish specific instances: the theoretical curves in Fig. \ref{Fig4} indicate that different entropies are dissipated for distinct initial pure states, and the lower bound \eqref{lower} reflects such behaviour. Our results confirm that the excited state $|H\rangle$ asymptotically demands more entropy dispersion to be erased via thermalization.

%

{\it Conclusions.} 
We have shown that the entropy production in open quantum systems can be bounded, both from above and below, with quantities that depend on purely geometrical features of the Hilbert space. Such bounds have been retrieved by explicitly comparing the Quantum Fisher metrics with the Wigner-Yanase one, representing the only two examples for which an analytic expression of the associated geodesic length is known. According to our analysis, supported by experimental evidences provided in a quantum optics experiment, the Wigner-Yanase metrics yields the tighter bound. Finally this study, based on the comparison between different metrics on the Hilbert space of the system, shows explicit connections with the computation of the geometrical quantum speed limits~\cite{Pires2016}, thus offering an interesting connection between entropy and time in irreversible physical phenomena.
 
{\it Acknowledgments.} We are grateful to Andrea Mari, Mario Arnolfo Ciampini, Roberto Raimondi, and Mauro Paternostro for insightful feedbacks on the manuscript.


%
%

\newpage

\widetext
\section*{Supplementary Information}
We prove the relation (\ref{tin}) for a dynamical map of the form  $\Phi_{t}(\rho_0) = (1-\lambda_t) \rho_0 + \lambda_t\rho_{eq}$
where $\lambda(t) \in [0,1]$ is an increasing function of time.
We recall that for a diagonal initial state
the Generalized Amplitude Damping (GAD) Channel gives $\rho(t) = (1-\eta_t) \rho_0 + \eta_t \rho_{eq}$
where $\rho_{eq}= p \ket{0}\bra{0} + (1-p) \ket{1}\bra{1}$ and $ 0\leq \eta_t \leq 1 $, thus this kind of evolution 
falls in the class of dynamical maps introduced above.
Computing the density matrix
at times $t_1$ and $t_2\geq t_1$, we obtain
\ba \label{comb} \rho(t_1) = \big( 1- \frac{\lambda_{t_1}}{\lambda_{t_2}} \big)\rho_0 + \frac{\lambda_{t_1}}{\lambda_{t_2}} \rho(t_2),\ea
with $0 \leq \lambda_{t_1}/\lambda_{t_2} \leq 1$.
Finally, plugging the equation (\ref{comb}) in the right hand side of the (\ref{tin}),  the proof follows
from the convexity of the relative entropy in each argument.
Indeed 

\ba S(\rho(t_1)||\rho(t_2)) = S(  \big( 1- \frac{\lambda_{t_1}}{\lambda_{t_2}} \big)\rho_0 + \frac{\lambda_{t_1}}{\lambda_{t_2}} \rho(t_2) || \rho(t_2)) \leq \big( 1-\frac{\lambda_{t_1}}{\lambda_{t_2}}\big) S(\rho_0 ||\rho(t_2)),\ea

\ba S(\rho_0||\rho(t_1)) = S(  \rho_0 ||\big( 1- \frac{\lambda_{t_1}}{\lambda_{t_2}} \big) \rho_0 +  \frac{\lambda_{t_1}}{\lambda_{t_2}} \rho(t_2)) \leq \frac{\lambda_{t_1}}{\lambda_{t_2}} S(\rho_0 ||\rho(t_2)),\ea

and the relation (\ref{tin}) is obtained summating the two equations above.
In the case of the GAD Channel this result has been confirmed and extended to the coherent case with a numerical procedure.
Furthermore, again restricting our attention to the GAD, it can be numerically shown that the sharpest bound~\eqref{upper} is retrieved by choosing the Wigner-Yanase metrics $X=WY$, as also confirmed by the experiment (see Supplemental Figure~\ref{SuppFig}). This shows the bound
\begin{equation}
S(\rho(t)||\rho_{eq}) \geq \frac{8}{\pi^2} \mathcal{L}_{X}(\rho(t),\rho_{eq}),
\end{equation}
for $X=WY$ and $X=QF$. The Wigner-Yanase bound always outperform the analogue based on the Bures metric for states with quantum coherence, such as the superposition $\vert D\rangle$, while the two metric give the same results for diagonal states in the energy eigenbasis, such as $\vert H \rangle$.

\begin{figure*}[h]
\includegraphics[width=0.49\columnwidth]{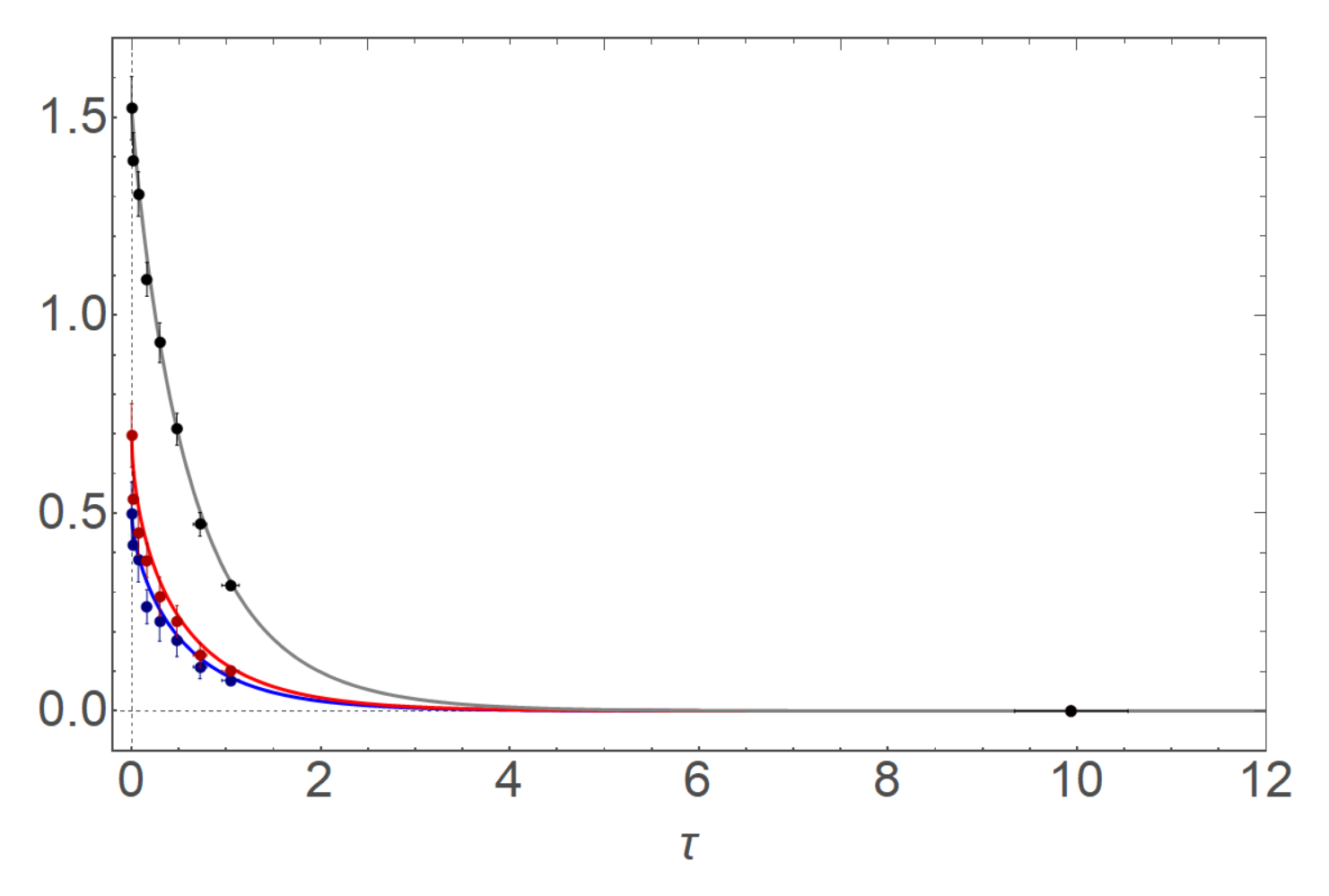}
\includegraphics[width=0.49\columnwidth]{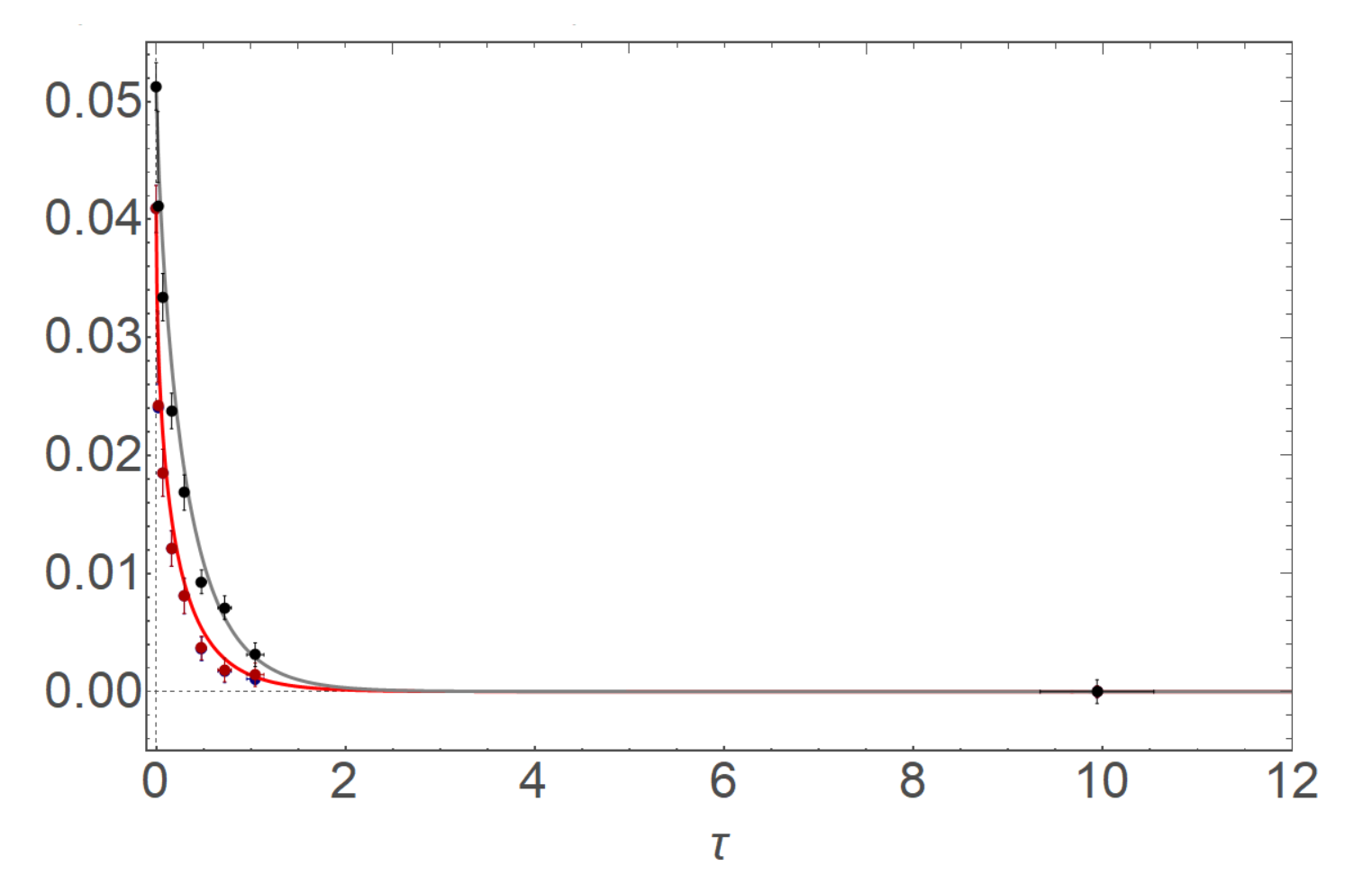}
\caption{ Geometric bound on $S(\rho(t)||\rho_{eq})$, the conditional entropy of the state $\rho(t)$ with respect the thermalised state $\rho_eq$. Left panel corresponds to the initial state $\vert D \rangle$, right panel to the intitial state $\vert V \rangle$. The points correspond to the experimental simulation, the solid lines to the theoretical predictions. Black points and solid line: conditional entropy; red points and solid line: WY metric; blue points and solid line: QF metric. Notice that the predictions for the WY and the QF metric are identical for the classical state $\vert V \rangle$}
\label{SuppFig}
\end{figure*}

\end{document}